\documentclass[letterpaper,10pt]{article} 

\usepackage{opticameet3} 

\usepackage{amsmath, amssymb, amsthm}
\usepackage{booktabs}
\usepackage{nicefrac}
\usepackage{microtype}
\usepackage{graphicx}
\usepackage{multirow}
\usepackage{float} 
\usepackage{tabularx}
\usepackage{subcaption}
\usepackage{makecell}
\usepackage{xspace}
\usepackage{soul}

\usepackage{tikz}
\usetikzlibrary{positioning}  
\usetikzlibrary{arrows.meta}  
\usetikzlibrary{shapes, shapes.geometric}  
\usetikzlibrary{fit}  

\usepackage{algorithm}
\usepackage{algpseudocode}

\usepackage[numbers]{natbib} 
\usepackage[colorlinks=true,bookmarks=false,citecolor=blue,urlcolor=blue]{hyperref}

\makeatletter
\renewcommand{\@biblabel}[1]{[#1]}
\makeatother

\newtheorem{theorem}{Theorem}

\newtheorem*{theorem*}{Theorem}
\newtheorem*{lemma*}{Lemma}
\newtheorem*{corollary*}{Corollary}

\newcommand\authormark[1]{\textsuperscript{#1}}
\newcommand{\ModelName}{\textsc{\textit{MTI}}~V.1\xspace}

\begin{document}

\title{Can Transformer Memory Be Corrupted? Investigating Cache-Side Vulnerabilities in Large Language Models}


\author{Elias Hossain,\authormark{1,*} Swayamjit Saha,\authormark{2} Somshubhra Roy,\authormark{3} and Ravi Prasad\authormark{4}}

\address{
\authormark{1}College of Engineering and Computer Science, University of Central Florida, Orlando, FL 32816, United States\\
\authormark{2,4}Department of Computer Science and Engineering, Mississippi State University, Mississippi State, MS 39762, United States\\
\authormark{3}Department of Electrical and Computer Engineering, North Carolina State University, Raleigh, NC 27695-7911, United States\\
}

\email{\authormark{*}Corresponding author: mdelias.hossain@ucf.edu} 

\begin{abstract}
Even when prompts and parameters are secured, transformer language models can remain exposed because their key--value (KV) cache during inference constitutes an under-examined attack surface. We present \ModelName, a modular framework that formalizes and implements Malicious Token Injection (MTI): cache-side perturbations applied to cached key vectors at selected layers and timesteps, with controllable magnitude and frequency (e.g., additive Gaussian noise, zeroing, and orthogonal rotations). We provide a theoretical analysis that upper-bounds how bounded cache perturbations propagate through attention, linking (i) logit deviations to the perturbation norm and query scaling, and (ii) shifts in attention and output distributions to these logit changes via standard softmax stability arguments under bounded-logit assumptions. Empirically, we show that \ModelName induces measurable distributional shifts and downstream performance degradation on GPT-2 and LLaMA-2/7B across synthetic prompts and standard NLP benchmarks. Extending to retrieval-augmented and agentic pipelines, we find that perturbations in post-retrieval layers can destabilize grounding and action selection. Collectively, these results motivate treating cache integrity as a first-class robustness boundary and position controlled KV perturbation as a reproducible experimental threat model for future research.
\end{abstract}

\section{Introduction}
\label{sec:intro}

Large language models (LLMs) have become foundational components of modern AI
systems, enabling applications such as retrieval-augmented generation (RAG),
autonomous agents, decision support, and long-horizon planning
\citep{brown2020language, touvron2023llama, team2024gemma}.
In deployed settings, their reliability depends not only on training-time alignment
and prompt-level safeguards, but also on the integrity of inference-time execution.
However, the robustness of internal inference-time state has received
comparatively limited attention in the trustworthy AI literature.

A key mechanism enabling efficient inference in decoder-only transformers is the
key--value (KV) cache, which stores intermediate attention representations across
decoding steps.
While essential for efficient inference, KV caching also introduces a persistent
internal state that directly influences future predictions and has not been widely
examined as a robustness boundary.

Cache-side perturbations differ fundamentally from prompt injection and
training-time attacks.
They operate entirely after tokenization and embedding, do not modify prompts or
model parameters, and therefore bypass input-level filters and alignment checks.
Even small, structured perturbations to cached representations can bias attention
distributions and propagate through downstream predictions without triggering
conventional safeguards.
This observation motivates a central question for trustworthy AI:
\emph{Can inference-time cache corruption systematically undermine reliability,
calibration, and grounding in modern LLM pipelines?}

To address this question, we introduce \ModelName, a modular framework for studying
cache-side perturbations during inference.
The framework formalizes MIT as controlled modifications to cached key vectors at
selected layers and timesteps, with tunable magnitude, frequency, and structure.
We provide a theoretical analysis that bounds how norm-limited cache perturbations
propagate through attention and affect next-token logits.
Empirically, we show that cache corruption induces distributional shift,
miscalibration, and task failure across GPT-2, LLaMA-2, and Gemma models, spanning
language modeling, extractive question answering, summarization,
retrieval-augmented generation, and agentic reasoning pipelines.

Our contributions are summarized as follows:
\begin{itemize}
    \item We identify inference-time KV cache integrity as a critical and
    underexplored robustness boundary, and formalize cache-side perturbation
    as a principled threat model for trustworthy AI evaluation.
    \item We introduce a modular framework for controlled cache perturbation
    and provide theoretical bounds characterizing how bounded cache corruption
    propagates to attention distributions and output logits.
    \item We empirically demonstrate that cache perturbations induce
    distributional shift, miscalibration, and grounding failures across
    language modeling, retrieval-augmented generation, and agentic reasoning
    pipelines.
    \item We evaluate lightweight inference-time cache defenses that partially
    mitigate corruption with minimal runtime overhead.
\end{itemize}

\section{Related Work}

Prior research on transformer robustness and security broadly spans architectural
efficiency, model integrity, and inference-time cache behavior.
On the architectural side, Multi-Head Latent Attention (MLA) has emerged as an
alternative to dense self-attention, introducing latent bottlenecks that reduce
memory bandwidth while preserving long-context performance.
DeepSeek-V2 first systematized MLA at scale \cite{liu2024deepseek}, with subsequent
iterations refining routing and optimization for improved efficiency and reasoning
capabilities \cite{liu2024deepseek2,guo2025deepseek}, and TransMLA providing a
formalized, research-oriented implementation \cite{meng2025transmla}.
In parallel, decoder-only Transformers universally rely on key--value (KV) caching
to accelerate autoregressive decoding, as exemplified by GPT-3 \cite{brown2020language},
LLaMA-2 \cite{touvron2023llama}, Mistral-7B \cite{jiang2023mistral7b}, Mixtral 8$\times$7B
\cite{jiang2024mixtral}, and the Qwen series \cite{bai2023qwen}.
While these systems emphasize throughput and scalability, the KV cache is typically
treated as an implementation detail rather than a robustness or security boundary.
These models are referenced to contextualize contemporary KV-cache usage and
optimization trends, rather than evaluated empirically.

A separate line of work investigates attacks on model parameters themselves,
including weight poisoning and backdoors in pretrained encoders \cite{li2021backdoor},
direct parameter editing to induce domain-specific harms in open LLMs
\cite{han2024medical}, hardware-level bit-flip attacks under realistic fault models
\cite{almalky2025vulnerable}, and trojan insertion in code and clinical language
models \cite{hussain2024trojans,lyu2025badclm}, with recent surveys synthesizing
these threat models and defenses \cite{zhao2024survey}.
More recently, attention has turned to the KV cache as a source of privacy and
security risk during inference, including cross-session leakage in multi-tenant
settings \cite{wu2025know}, timing-based side channels \cite{song2024early}, prompt
reconstruction attacks and cache obfuscation defenses \cite{luo2025shadow}, and
restricted cache reuse policies \cite{chu2025selective}.
However, existing cache-centric studies primarily focus on passive leakage rather
than active manipulation.
In contrast, this work treats the KV cache as a first-class adversarial surface and
systematically studies how structured perturbations to cached keys induce
distributional shift, miscalibration, and downstream failure during inference,
bridging a critical gap between efficiency-driven caching mechanisms and
robustness-oriented security analysis.

\section{Cache Model and Threat Assumptions}
\label{sec:prelim}

Transformer decoders maintain a persistent \emph{KV cache} during
autoregressive decoding, storing projected representations of previously generated
tokens to avoid recomputing attention over the full history.
While essential for efficient inference, this cache also introduces a stateful
internal memory that mediates future predictions.
We treat the KV cache as a distinct inference-time robustness boundary and formalize
the associated threat model below.

\paragraph{KV Cache Mechanism.}
At decoding step $t$, each layer projects token embeddings into queries, keys, and
values, and computes attention via scaled dot products.
Previously computed keys and values $\{k_j, v_j\}_{j=1}^{t-1}$ are stored in a cache
and reused at subsequent steps, making cached representations persistent inputs
to future attention computations.

\paragraph{Threat Model.}
We consider an adversary with inference-time access to the KV cache.
The adversary cannot modify prompts or model parameters, but can perturb cached keys
at selected layers and timesteps.
Formally, at layer $\ell$ and position $j$,
\[
k^{(\ell)}_j \mapsto \tilde{k}^{(\ell)}_j = k^{(\ell)}_j + \delta^{(\ell)}_j ,
\]
where $\delta^{(\ell)}_j$ is a corruption vector.
Perturbations may be stochastic (e.g., Gaussian noise), destructive (e.g., zeroing),
or structured (e.g., orthogonal rotations), and are controlled in magnitude and
injection frequency.
Unless stated otherwise, our theoretical analysis applies to non-adaptive,
norm-bounded perturbations; optimized variants are evaluated empirically.

\paragraph{Scope and Relation to Prior Threat Models.}
This threat model captures settings where cache state can be modified during
inference without altering inputs or weights, such as compromised runtimes or
misconfigured shared deployments.
Cache perturbations differ from prompt injection, which operates at the input
interface, and from weight poisoning, which requires training-time access.
Unlike cache leakage or side-channel attacks, which are passive, we study
\emph{active} manipulation of inference-time state.
Accordingly, cache corruption represents a complementary robustness threat within
the broader trustworthy AI landscape.

\section{\ModelName Framework}
\label{sec:attack}

We formalize \emph{Malicious Token Injection (MTI)}, i.e., cache-side perturbations
that modify stored attention key representations during inference without altering
input tokens, prompts, or model parameters.
The central observation underlying \ModelName is that attention weights are
determined by inner products between queries and cached keys; consequently, even
small but structured modifications to cached keys can bias attention
distributions and alter downstream token predictions. Rather than defining a single attack, \ModelName specifies a family of cache
perturbations controlled along three primary axes:
(i) corruption magnitude, which determines the strength of the injected
perturbation;
(ii) injection frequency, which governs how often perturbations are applied across
decoding steps; and
(iii) layer placement, which specifies which transformer layers are targeted.
This parameterization enables systematic exploration of cache sensitivity without
assuming adversarial optimality.

\paragraph{Perturbation Classes.}
Within this framework, we study three canonical corruption mechanisms that span
distinct modes of cache integrity violation.
\textbf{MTI-Gaussian} injects additive noise sampled from
$\mathcal{N}(0,\sigma^2 I)$ into cached keys, modeling stochastic corruption.
\textbf{MTI-Zeroing} removes cached keys entirely, simulating catastrophic cache
erasure and representing an extreme but informative failure mode.
\textbf{MTI-Rotation} applies orthogonal transformations to cached keys, inducing
structured yet norm-preserving perturbations that alter angular alignment with
queries while preserving vector magnitude.
Together, these mechanisms capture random noise, deletion, and structured
misalignment, which represent three fundamental manifestations of cache
perturbation during transformer inference.

\paragraph{Optimized Cache Perturbation.}
In addition to fixed-form perturbations, we also consider an optional
optimization-based variant that computes cache perturbations designed to maximize
an adversarial objective at inference time.
This variant is used to characterize worst-case empirical behavior and is not
assumed by the theoretical analysis.
Specifically, Algorithm~\ref{alg:mti_opt} describes an inner-loop procedure that
iteratively updates a perturbation applied to a selected cached key by ascending
the gradient of an adversarial loss, subject to a norm constraint.
The procedure operates entirely on cached key representations and does not modify
model weights, input prompts, or value vectors.
All optimized perturbations are evaluated empirically and fall outside the scope
of the formal bounds in Section~\ref{sec:theory}, which apply exclusively to
non-adaptive, norm-bounded perturbations.

\begin{algorithm}[!htb]
\caption{OptimizePerturbation: Inner-loop adversarial perturbation}
\label{alg:mti_opt}
\begin{algorithmic}[1]
\Require model $\mathcal{M}$, current caches $\mathcal{C}$, query
$q_t^{(\ell,h)}$, target position $j$, optimizer config $\mathcal{O}$
(steps $K$, step size $\eta$, loss $\mathcal{L}_{\text{adv}}$), norm constraint
$\epsilon$
\State initialize perturbation $\delta \leftarrow \mathbf{0}$ (or small random noise)
\For{$k \leftarrow 1$ \textbf{to} $K$}
    \State compute tentative key $\tilde{k} \leftarrow k_j + \delta$
    \State update cache copy $\mathcal{C}'$ with $\mathcal{C}'[j]\leftarrow \tilde{k}$
    \State evaluate adversarial loss
    $\ell \leftarrow \mathcal{L}_{\text{adv}}(\mathcal{M},\mathcal{C}', q_t)$
    \State compute gradient $g \leftarrow \nabla_\delta \ell$
    \State update
    $\delta \leftarrow \delta + \eta \cdot
    \mathrm{proj}_{\|\cdot\|\le \epsilon}(g)$
\EndFor
\State \Return optimized perturbation $\delta$
\end{algorithmic}
\end{algorithm}

\section{Theoretical Analysis of Cache Perturbations}
\label{sec:theory}

We analyze how cache-side perturbations propagate through transformer attention
and influence the logits used for next-token prediction.
Because attention scores are computed via inner products $q_t^\top k_j$, any
modification to cached keys directly alters attention weights and, in turn, the
aggregated value representations.
Our goal is to characterize the \emph{worst-case sensitivity} of attention and
logits to bounded cache corruption, rather than to provide tight estimates of
typical inference-time behavior.

\paragraph{Perturbation-to-Logit Bound.}
Let $\delta_j$ denote a perturbation applied to a cached key $k_j$ at timestep $j$,
with $\|\delta_j\|_2 \le \epsilon$.
The resulting deviation in the attention score
$q_t^\top (k_j + \delta_j)$ relative to the clean score scales linearly with both
the perturbation magnitude $\epsilon$ and the query norm $\|q_t\|_2$.
Aggregating across positions yields a worst-case bound on the deviation of the
unnormalized logits:
\[
\|\Delta z_t\|_2 \le \|q_t\|_2 \cdot \epsilon,
\]
where $\Delta z_t$ denotes the logit deviation induced by cache perturbation.
For clarity, normalization constants (e.g., the $\sqrt{d}$ factor in scaled
dot-product attention) are absorbed into $\epsilon$, as our analysis focuses on
asymptotic scaling behavior rather than tight constant factors.

\paragraph{Softmax Stability.}
To relate logit deviations to downstream effects, we appeal to standard softmax
stability results under bounded-logit assumptions.
Provided that logits remain within a bounded range, as is typical under
temperature scaling and finite-precision inference, the softmax mapping is
Lipschitz continuous on this restricted domain:
\[
\|\mathrm{softmax}(z + \Delta z) - \mathrm{softmax}(z)\|_1
\le L \cdot \|\Delta z\|_2,
\]
where $L$ depends on the bound of the logit domain.
This property allows deviations in logits to be translated into controlled
changes in the resulting attention distribution.

\paragraph{Theorem.}
\emph{For any transformer layer $\ell$, if an adversary injects perturbations
$\{\delta_j\}$ into cached keys with total norm bounded by $\epsilon$, then the
resulting deviation in the next-token logits is bounded by
$O(\epsilon \cdot \|q_t\|_2)$.
Consequently, the induced change in the attention distribution is Lipschitz-bounded
to the same order.}
Formal proofs and extensions to multi-head attention are provided in
Appendix~\ref{app:proofs}.

\paragraph{Interpretation and Empirical Context.}
These bounds characterize \emph{worst-case scaling behavior} under norm-bounded,
non-adaptive cache perturbations and are not intended to tightly predict typical
inference-time deviations.
Empirically, measured logit deviations under \ModelName perturbations remain well
below the corresponding theoretical upper bounds across all evaluated models and
layers.
This gap indicates that the analysis is deliberately conservative, serving to
formalize sensitivity trends rather than to provide tight performance guarantees.

\section{Setup}
\label{app:exp-setup}

We evaluate \ModelName across a diverse set of benchmarks spanning short-sequence
classification, extractive question answering, long-sequence summarization,
multi-hop reasoning, and retrieval-augmented generation.
Sentiment classification uses SST-2 \cite{socher-etal-2013-recursive}, question
answering uses SQuAD v1.1 \cite{rajpurkar-etal-2016-squad} and HotpotQA
\cite{yang2018hotpotqadatasetdiverseexplainable}, and summarization is evaluated on
CNN/DailyMail \cite{see-etal-2017-get}.
Retrieval-augmented experiments use Wikipedia passages from the 2021 dump, where we
report grounding and hallucination rates in factual QA.
All datasets follow standard splits and are tokenized using HuggingFace tokenizers,
with inputs truncated to model-specific context limits (512 tokens for GPT-2 and up
to 4k tokens for LLaMA-2 and Gemma).

We report both distributional and task-level metrics to capture the effects of cache
perturbations.
Distributional shift is quantified using KL divergence and perplexity change,
reflecting changes in confidence and calibration.
Task performance is measured using accuracy for SST-2, F1 and Exact Match for
question answering, and ROUGE-L for summarization.
Perturbation strength is varied systematically using Gaussian noise
($\sigma \in \{0.01, 0.05, 0.1, 0.2\}$), rotation angles
($\theta \in \{15^\circ, 30^\circ, 45^\circ\}$), and zeroing ratios
($\{0.5\%, 1\%, 5\%\}$).
All results are averaged over three random seeds, with statistical significance
verified using paired $t$-tests ($p<0.05$).

Experiments are conducted using GPT-2 Medium \cite{radford2019language},
LLaMA-2 7B \cite{touvron2023llama}, and Gemma-7B \cite{team2024gemma}, representing
different architectural scales.
All models are evaluated in inference-only mode using HuggingFace Transformers
(v4.39) and official checkpoints.
Experiments run on NVIDIA T4 (16GB) or A100 (40GB) GPUs with mixed precision
(\texttt{float16}) and batch size 16.
Cache perturbations are injected directly into stored key vectors at selected
layers, typically the 6th layer for GPT-2 and the 12th layer for LLaMA-2 and Gemma,
using either continuous or intermittent injection schedules.


\section{Experimental Results}
\label{sec:experiments}

This section presents an empirical evaluation of \ModelName across token-level
language modeling behavior, downstream NLP tasks, retrieval-augmented generation,
agentic reasoning pipelines, and runtime scalability.
Extended results and ablation studies are provided in
Appendix~\ref{app:extended-results}.

\subsection{Token-Level Distributional Effects}
\label{sec:token-main}

We analyze the impact of \ModelName on next-token probability distributions in
standard language modeling settings.
Rather than focusing solely on downstream task accuracy, this analysis probes the
internal probabilistic structure of model outputs.
Table~\ref{tab:token-main} reports representative results across GPT-2 and
LLaMA-2 models using KL divergence relative to clean runs, Top-1 accuracy drop,
and perplexity change.
Across all evaluated settings, cache perturbations induce KL divergences that are
an order of magnitude larger than stochastic variation observed under different
random seeds, indicating \emph{structural rather than incidental} effects. Larger KL divergence correlates strongly with task-level degradation.
For example, GPT-2 Medium under rotation-based corruption exhibits the largest
distributional shift (KL $=0.57$), accompanied by a 16.7\% Top-1 accuracy drop.
Perplexity exhibits a nuanced pattern: Gaussian perturbations increase uncertainty,
while zeroing collapses distributions into degenerate low-entropy regimes,
reflecting overconfidence rather than improved modeling.

\begin{table}[t]
\centering
\caption{Representative token-level effects under \ModelName.
KL divergence is measured relative to clean runs.}
\label{tab:token-main}
\begin{tabular}{l l c c c}
\toprule
\textbf{Model} & \textbf{Attack} & \textbf{KL} & \textbf{Acc Drop} & \textbf{PPL} \\
\midrule
GPT-2 Medium & Rotation & 0.57 & 16.7\% & +130.6\% \\
LLaMA-2/7B  & Zeroing  & 0.20 & 0.0\%  & --43.1\% \\
\bottomrule
\end{tabular}
\end{table}

\subsection{Analyzing Defense Effectiveness}
\label{app:defense}

We evaluate whether lightweight cache-level defenses can mitigate \ModelName with
minimal computational overhead (Table~\ref{tab:defense}).
We consider three strategies: \emph{Cache Reset}, which periodically clears cached
entries at a target layer; \emph{Dropout Mask Randomization}, which stochastically
masks subsets of key--value vectors; and \emph{Attention Smoothing}, which applies
temporal averaging to dampen abrupt perturbations.
Evaluation is performed on the MRPC task (GLUE), using accuracy as the primary
metric and runtime overhead measured relative to clean inference.
Overhead below $1.0\times$ reflects reduced cache growth due to periodic reset,
which can slightly reduce attention computation. Across GPT-2 Medium and LLaMA-2 7B, all three defenses preserve baseline accuracy
under mild cache perturbations while incurring minimal runtime cost
($0.69\times$ to $1.08\times$).

Identical accuracies across defenses reflect the mild perturbation regime used for
evaluation, in which defenses primarily prevent further degradation rather than
recover lost performance.
Gemma-7B exhibits $0.0\%$ accuracy even in the clean baseline due to
tokenizer–label misalignment in this setup; we therefore report its results for
completeness but exclude them from aggregate conclusions. Overall, these results suggest that cache-level defenses offer limited but
meaningful protection against mild cache corruption.
However, they are insufficient as standalone safeguards and should be viewed as
preliminary measures rather than comprehensive solutions, motivating future work
on stronger cache integrity mechanisms and inference-time monitoring.

\begin{table}[H]
\centering
\caption{Defense effectiveness against \ModelName under mild cache perturbations.
Accuracy is reported under attack with the specified defense.
Gemma-7B results are shown for completeness but excluded from aggregate conclusions
due to tokenizer–label misalignment in this task.}
\label{tab:defense}
\begin{tabular}{l l l c c c}
\toprule
\textbf{Model} & \textbf{Defense Method} & \textbf{Attack Type} & 
\textbf{Accuracy (\%)} & \textbf{Degradation} & \textbf{Overhead} \\
\midrule
LLaMA-2 7B & None (Baseline) & Zeroing  & 60.0 & 0.0\% & 1.00$\times$ \\
LLaMA-2 7B & Cache Reset     & Zeroing  & 60.0 & 0.0\% & 0.69$\times$ \\
LLaMA-2 7B & Dropout Mask Rand. & Gaussian & 60.0 & 0.0\% & 0.70$\times$ \\
LLaMA-2 7B & Attention Smoothing & Rotation & 60.0 & 0.0\% & 0.75$\times$ \\
Gemma-7B   & None (Baseline) & Zeroing  &  0.0 & 0.0\% & 1.00$\times$ \\
Gemma-7B   & Cache Reset     & Zeroing  &  0.0 & 0.0\% & 1.00$\times$ \\
Gemma-7B   & Dropout Mask Rand. & Gaussian &  0.0 & 0.0\% & 1.00$\times$ \\
Gemma-7B   & Attention Smoothing & Rotation &  0.0 & 0.0\% & 1.00$\times$ \\
GPT-2 Medium & None (Baseline) & Zeroing  & 60.0 & 0.0\% & 1.00$\times$ \\
GPT-2 Medium & Cache Reset     & Zeroing  & 60.0 & 0.0\% & 1.00$\times$ \\
GPT-2 Medium & Dropout Mask Rand. & Gaussian & 60.0 & 0.0\% & 1.01$\times$ \\
GPT-2 Medium & Attention Smoothing & Rotation & 60.0 & 0.0\% & 1.08$\times$ \\
\bottomrule
\end{tabular}
\end{table}

\subsection{Downstream Tasks, Retrieval, and Agentic Pipelines}
\label{sec:systems-main}

\begin{figure}[t]
    \centering
    \includegraphics[width=0.9\linewidth]{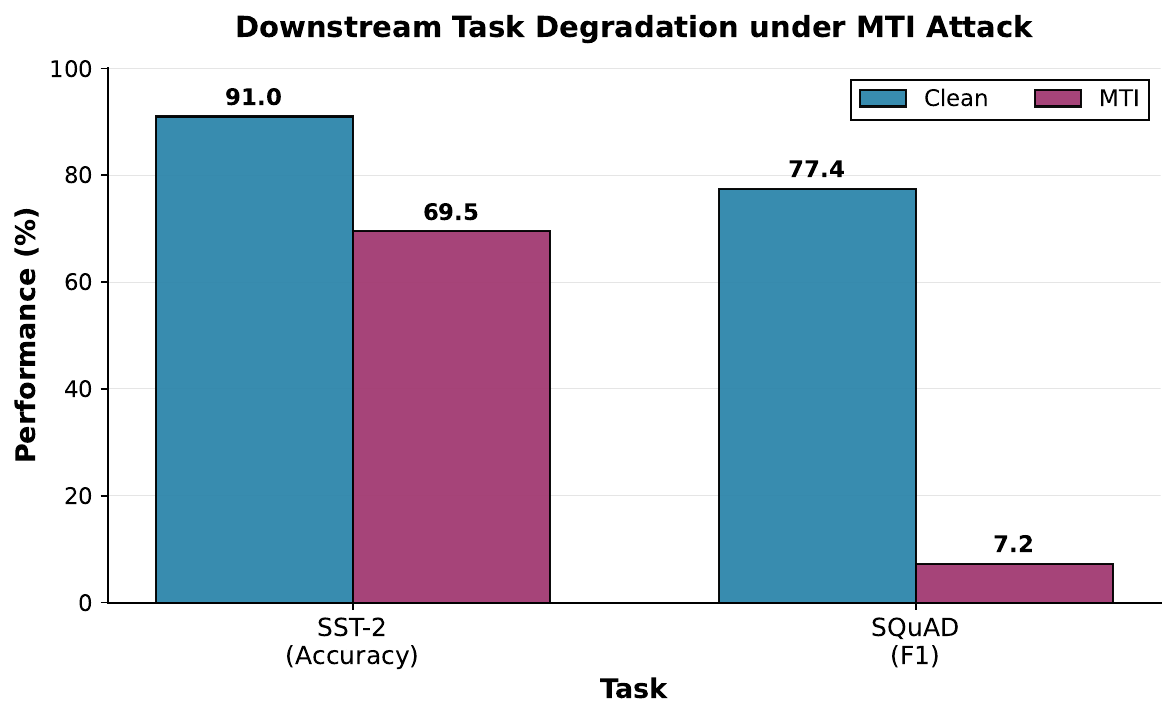}
    \caption{Downstream task degradation under \ModelName.
    Performance on SST-2 (accuracy) and SQuAD v1.1 (F1) is shown for clean inference
    and under cache perturbation (MTI).
    Span-based reasoning tasks exhibit significantly higher sensitivity to cache
    corruption.}

    \label{fig:downstream-degradation}
    \vspace{-4mm}
\end{figure}

We evaluate the downstream impact of cache perturbations across standard NLP
tasks, retrieval-augmented generation (RAG), and agentic reasoning pipelines,
highlighting how cache corruption manifests under different system structures. For standard NLP tasks, we evaluate SST-2 sentiment classification and SQuAD v1.1
extractive question answering.
As shown in Figure~\ref{fig:downstream-degradation}, SST-2 exhibits moderate
robustness, with strong perturbations reducing accuracy by over 23\%.
In contrast, SQuAD exhibits catastrophic failure under identical perturbations,
with F1 scores collapsing from 77.4 to 7.2. This divergence indicates that tasks requiring precise token alignment and
long-context reasoning are particularly sensitive to cache corruption.

We next evaluate retrieval-augmented generation on HotpotQA.
Perturbations applied prior to retrieval have limited effect, whereas perturbations
applied to retrieved context significantly degrade grounding fidelity and increase
hallucination rates.
These results indicate that RAG pipelines depend critically on cache integrity at
the evidence stage.
Detailed results are provided in Appendix~\ref{app:rag}. Finally, we evaluate \ModelName in agentic reasoning frameworks, including ReAct
and AutoGPT-style planning.
Success rates remain unchanged under mild cache perturbations, suggesting that
closed-loop reasoning can partially buffer localized corruption.
However, this robustness should be interpreted cautiously, as perturbations are
limited to mild regimes.
Extended results are reported in Appendix~\ref{app:agents}.


\section{Ablation Studies}
\label{app:ablations}

\noindent\textbf{Model choice.}
We use DistilBERT for ablation analysis due to its reduced depth, which enables
systematic sweeps over layer placement, perturbation magnitude, and injection
frequency while preserving standard transformer attention.
Although DistilBERT is encoder-only, cache perturbations are emulated by injecting
corruption into stored attention keys, enabling controlled analysis of key
sensitivity independent of autoregressive decoding.

These ablations are intended to capture qualitative trends rather than
architecture-specific vulnerability. We analyze the effects of layer placement, perturbation magnitude, and injection
frequency on cache-side perturbations.
Layer-wise analysis shows that mid-level layers are consistently more vulnerable
than early or deep layers: KL divergence increases from $0.0367$ at Layer~1 to
$0.0447$ at Layer~3 under identical perturbation budgets, while deeper layers
exhibit negligible additional sensitivity.

This pattern suggests that mid-level representations, which mediate semantic
integration, constitute a critical bottleneck for cache integrity. We further ablate perturbation magnitude at a fixed layer.
As shown in Table~\ref{tab:ablation-magnitude}, Gaussian perturbations exhibit a
non-monotonic effect, with intermediate noise producing the largest degradation,
while extreme noise collapses representations and limits further drift.
Zeroing induces small but persistent degradation, whereas rotation-based
perturbations remain largely scale-invariant due to their norm-preserving nature.

Injection frequency further modulates attack effectiveness.
Continuous corruption produces substantially greater degradation than
intermittent schedules, indicating that cache perturbations are stateful and
compound over time.
Joint ablations over layer depth and perturbation magnitude
(Table~\ref{tab:combined}) reveal strong interaction effects in DistilBERT, where
mid-layer perturbations become increasingly destabilizing as magnitude increases,
while GPT-2 Medium remains largely invariant across both dimensions. Overall, these ablations show that cache vulnerability is governed by the
interaction of layer placement, perturbation strength, and temporal persistence,
rather than any single factor in isolation.

\begin{table}[!t]
\centering
\caption{Ablation results over perturbation magnitude and layer placement.
Values denote relative accuracy degradation.}
\label{tab:ablation-magnitude}
\begin{tabular}{l c c c c}
\toprule
\textbf{Setting} & \textbf{Low} & \textbf{Medium} & \textbf{High} & \textbf{Extreme} \\
\midrule
Gaussian (DistilBERT L3) & -0.0092 & 0.0195 & -0.0264 & -0.0103 \\
Rotation (DistilBERT L3) & -0.0023 & 0.0000 & 0.0000 & 0.0000 \\
Zeroing (DistilBERT L3)  & -0.0023 & 0.0057 & 0.0023 & 0.0011 \\
\midrule
DistilBERT Layer 1 & 0.330 & 0.385 & 0.385 & -- \\
DistilBERT Layer 3 & 0.225 & 0.400 & 0.395 & -- \\
GPT-2 Medium Layer 6 & 0.000 & 0.000 & -0.055 & -- \\
\bottomrule
\end{tabular}
\end{table}

\begin{table}[H]
\centering
\caption{Joint ablation across layer choice and perturbation magnitude. Reported
values indicate relative degradation in accuracy.}
\label{tab:combined}
\begin{tabular}{l c c c c}
\toprule
\textbf{Model} & \textbf{Layer} & \textbf{Low} & \textbf{Medium} & \textbf{High} \\
\midrule
DistilBERT-SST2 & 1 & 0.330 & 0.385 & 0.385 \\
DistilBERT-SST2 & 3 & 0.225 & 0.400 & 0.395 \\
GPT-2 Medium    & 1 & 0.000 & 0.000 & 0.000 \\
GPT-2 Medium    & 3 & 0.000 & 0.000 & 0.000 \\
GPT-2 Medium    & 6 & 0.000 & 0.000 & -0.055 \\
\bottomrule
\end{tabular}
\end{table}

\section{Discussion}
\label{sec:discussion}

Our results demonstrate that cache-side perturbations can induce meaningful
distributional shifts across a range of transformer-based systems.
Even lightweight and structured perturbations are sufficient to bias attention
dynamics and degrade downstream task performance under the assumed cache-access
threat model.
Importantly, these effects extend beyond single-turn prediction to more complex
pipelines, including retrieval-augmented generation and multi-step agentic
reasoning, where cache corruption can destabilize grounding and planning
consistency. The computational overhead associated with cache perturbation is minimal,
suggesting feasibility when such cache access exists.
However, whether this access is available in practice depends critically on
deployment assumptions, including isolation guarantees, serving infrastructure,
and systems-layer security.
Accordingly, our findings should be interpreted as characterizing a robustness
boundary rather than asserting universal exploitability across all deployments.

While simple cache-level defenses provide partial mitigation, none fully eliminate
the effects of persistent corruption.
This indicates that ad hoc countermeasures may be insufficient, and that explicit
cache integrity mechanisms or monitoring strategies may be necessary for robust
deployment.
Our analysis is limited by controlled perturbation regimes, selected datasets,
and the assumption of partial white-box cache access.
Extending this framework to long-context, multimodal, and adaptive systems
remains an important direction for future work. Beyond robustness and safety, cache-side perturbations also offer a controlled
probe into attention sensitivity.
Because \ModelName operates directly on attention keys, it provides a mechanism
for analyzing how small internal representation shifts alter attention allocation
and downstream behavior.
This suggests a potential connection between cache integrity and mechanistic
interpretability, where cache perturbations can be used not only as a threat
model but also as a diagnostic tool for understanding internal model dynamics.

\section{Conclusion}
\label{sec:conclusion}

We introduced \ModelName, a framework for analyzing cache-side perturbations that
target the internal key--value memory of transformer language models during
inference.
Across classification, question answering, retrieval-augmented generation, and
agentic pipelines, we demonstrated that inference-time cache corruption can
compromise reliability, calibration, and grounding, even when prompts and model
weights remain unchanged.
Retrieval-augmented systems exhibited pronounced sensitivity at the evidence
representation stage, while closed-loop agentic frameworks showed partial
resilience due to iterative feedback. We further evaluated lightweight cache-level defenses that preserve clean-task
accuracy with minimal runtime overhead.

While these defenses partially mitigate mild corruption, none provide robust
protection under stronger or persistent perturbations.
Ablation studies reveal that cache vulnerability arises from the interaction of
layer placement, perturbation magnitude, and temporal persistence, rather than
from any single factor alone. Taken together, these findings position inference-time cache integrity as a
first-class robustness concern for trustworthy AI.
By formalizing cache-side perturbation as a principled threat model and providing
both theoretical and empirical analysis, \ModelName establishes a foundation for
future work on inference-time monitoring, certified cache integrity, and
safety-aware deployment of large language models.

\paragraph{Limitations and Future Work}
\label{sec:limitations}

This work studies cache-side perturbations under a controlled inference-time threat
model and has several limitations.
First, we assume partial access to the key--value cache during inference, which may
not hold in fully isolated production deployments, but is realistic in compromised
runtimes or shared-serving environments. Second, our theoretical analysis focuses on non-adaptive, norm-bounded cache
perturbations.
While we empirically evaluate optimized variants, extending formal guarantees to
adaptive or sequential attacks remains an open problem. Third, our empirical evaluation is limited to a subset of transformer
architectures.
Future work will explore cache integrity in more recent and architecturally
distinct models, including systems with latent attention mechanisms,
mixture-of-experts routing, and extended context designs. Finally, the evaluated defenses are lightweight and exploratory.
Designing principled cache integrity mechanisms and certified inference-time
monitoring remains an important direction for future research.

\paragraph{Ethical Considerations.}
This work is intended to support robustness analysis and the design of trustworthy
AI systems.
We study cache perturbation as an explicit threat model under controlled
assumptions and do not release end-to-end exploitation tools.
The assumed cache-access capability may not hold in well-isolated deployments;
our goal is not to claim universal exploitability, but to surface a previously
overlooked inference-time failure mode and motivate defensive research on cache
integrity, monitoring, and safety-aware system design.

\bibliography{references}

\begin{thebibliography}{10}

\bibitem{brown2020language}
T.~Brown, B.~Mann, N.~Ryder, M.~Subbiah, J.~D. Kaplan, P.~Dhariwal, A.~Neelakantan, P.~Shyam, G.~Sastry, A.~Askell, {\em et~al.}, ``Language models are few-shot learners,'' {\em Advances in neural information processing systems}, vol.~33, pp.~1877--1901, 2020.

\bibitem{touvron2023llama}
H.~Touvron, T.~Lavril, G.~Izacard, X.~Martinet, M.-A. Lachaux, T.~Lacroix, B.~Rozi{\`e}re, N.~Goyal, E.~Hambro, F.~Azhar, {\em et~al.}, ``Llama: Open and efficient foundation language models,'' {\em arXiv preprint arXiv:2302.13971}, 2023.

\bibitem{team2024gemma}
G.~Team, T.~Mesnard, C.~Hardin, R.~Dadashi, S.~Bhupatiraju, S.~Pathak, L.~Sifre, M.~Rivi{\`e}re, M.~S. Kale, J.~Love, {\em et~al.}, ``Gemma: Open models based on gemini research and technology,'' {\em arXiv preprint arXiv:2403.08295}, 2024.

\bibitem{liu2024deepseek}
A.~Liu, B.~Feng, B.~Wang, B.~Wang, B.~Liu, C.~Zhao, C.~Dengr, C.~Ruan, D.~Dai, D.~Guo, {\em et~al.}, ``Deepseek-v2: A strong, economical, and efficient mixture-of-experts language model,'' {\em arXiv preprint arXiv:2405.04434}, 2024.

\bibitem{liu2024deepseek2}
A.~Liu, B.~Feng, B.~Xue, B.~Wang, B.~Wu, C.~Lu, C.~Zhao, C.~Deng, C.~Zhang, C.~Ruan, {\em et~al.}, ``Deepseek-v3 technical report,'' {\em arXiv preprint arXiv:2412.19437}, 2024.

\bibitem{guo2025deepseek}
D.~Guo, D.~Yang, H.~Zhang, J.~Song, R.~Zhang, R.~Xu, Q.~Zhu, S.~Ma, P.~Wang, X.~Bi, {\em et~al.}, ``Deepseek-r1: Incentivizing reasoning capability in llms via reinforcement learning,'' {\em arXiv preprint arXiv:2501.12948}, 2025.

\bibitem{meng2025transmla}
F.~Meng, P.~Tang, X.~Tang, Z.~Yao, X.~Sun, and M.~Zhang, ``Transmla: Multi-head latent attention is all you need,'' {\em arXiv preprint arXiv:2502.07864}, 2025.

\bibitem{jiang2023mistral7b}
A.~Q. Jiang, A.~Sablayrolles, A.~Mensch, C.~Bamford, D.~S. Chaplot, D.~de~las Casas, F.~Bressand, G.~Lengyel, G.~Lample, L.~Saulnier, L.~R. Lavaud, M.-A. Lachaux, P.~Stock, T.~L. Scao, T.~Lavril, T.~Wang, T.~Lacroix, and W.~E. Sayed, ``Mistral 7b,'' 2023.

\bibitem{jiang2024mixtral}
A.~Q. Jiang, A.~Sablayrolles, A.~Roux, A.~Mensch, B.~Savary, C.~Bamford, D.~S. Chaplot, D.~d.~l. Casas, E.~B. Hanna, F.~Bressand, {\em et~al.}, ``Mixtral of experts,'' {\em arXiv preprint arXiv:2401.04088}, 2024.

\bibitem{bai2023qwen}
J.~Bai, S.~Bai, Y.~Chu, Z.~Cui, K.~Dang, X.~Deng, Y.~Fan, W.~Ge, Y.~Han, F.~Huang, {\em et~al.}, ``Qwen technical report,'' {\em arXiv preprint arXiv:2309.16609}, 2023.

\bibitem{li2021backdoor}
L.~Li, D.~Song, X.~Li, J.~Zeng, R.~Ma, and X.~Qiu, ``Backdoor attacks on pre-trained models by layerwise weight poisoning,'' {\em arXiv preprint arXiv:2108.13888}, 2021.

\bibitem{han2024medical}
T.~Han, S.~Nebelung, F.~Khader, T.~Wang, G.~M{\"u}ller-Franzes, C.~Kuhl, S.~F{\"o}rsch, J.~Kleesiek, C.~Haarburger, K.~K. Bressem, {\em et~al.}, ``Medical large language models are susceptible to targeted misinformation attacks,'' {\em NPJ digital medicine}, vol.~7, no.~1, p.~288, 2024.

\bibitem{almalky2025vulnerable}
A.~M.~A. Almalky, R.~Zhou, S.~Angizi, and A.~S. Rakin, ``How vulnerable are large language models (llms) against adversarial bit-flip attacks?,'' in {\em Proceedings of the Great Lakes Symposium on VLSI 2025}, pp.~534--539, 2025.

\bibitem{hussain2024trojans}
A.~Hussain, M.~R.~I. Rabin, T.~Ahmed, B.~Xu, P.~Devanbu, and M.~A. Alipour, ``Trojans in large language models of code: A critical review through a trigger-based taxonomy,'' {\em arXiv preprint arXiv:2405.02828}, 2024.

\bibitem{lyu2025badclm}
W.~Lyu, Z.~Bi, F.~Wang, and C.~Chen, ``Badclm: Backdoor attack in clinical language models for electronic health records,'' in {\em AMIA Annual Symposium Proceedings}, vol.~2024, p.~768, 2025.

\bibitem{zhao2024survey}
S.~Zhao, M.~Jia, Z.~Guo, L.~Gan, X.~Xu, X.~Wu, J.~Fu, Y.~Feng, F.~Pan, and L.~A. Tuan, ``A survey of backdoor attacks and defenses on large language models: Implications for security measures,'' {\em Authorea Preprints}, 2024.

\bibitem{wu2025know}
G.~Wu, Z.~Zhang, Y.~Zhang, W.~Wang, J.~Niu, Y.~Wu, and Y.~Zhang, ``I know what you asked: Prompt leakage via kv-cache sharing in multi-tenant llm serving,'' in {\em Proceedings of the 2025 Network and Distributed System Security (NDSS) Symposium. San Diego, CA, USA}, 2025.

\bibitem{song2024early}
L.~Song, Z.~Pang, W.~Wang, Z.~Wang, X.~Wang, H.~Chen, W.~Song, Y.~Jin, D.~Meng, and R.~Hou, ``The early bird catches the leak: Unveiling timing side channels in llm serving systems,'' {\em arXiv preprint arXiv:2409.20002}, 2024.

\bibitem{luo2025shadow}
Z.~Luo, S.~Shao, S.~Zhang, L.~Zhou, Y.~Hu, C.~Zhao, Z.~Liu, and Z.~Qin, ``Shadow in the cache: Unveiling and mitigating privacy risks of kv-cache in llm inference,'' {\em arXiv preprint arXiv:2508.09442}, 2025.

\bibitem{chu2025selective}
K.~Chu, Z.~Lin, D.~Xiang, Z.~Shen, J.~Su, C.~Chu, Y.~Yang, W.~Zhang, W.~Wu, and W.~Zhang, ``Selective kv-cache sharing to mitigate timing side-channels in llm inference,'' {\em arXiv preprint arXiv:2508.08438}, 2025.

\bibitem{socher-etal-2013-recursive}
R.~Socher, A.~Perelygin, J.~Wu, J.~Chuang, C.~D. Manning, A.~Ng, and C.~Potts, ``Recursive deep models for semantic compositionality over a sentiment treebank,'' in {\em Proceedings of the 2013 Conference on Empirical Methods in Natural Language Processing}, (Seattle, Washington, USA), pp.~1631--1642, Association for Computational Linguistics, Oct. 2013.

\bibitem{rajpurkar-etal-2016-squad}
P.~Rajpurkar, J.~Zhang, K.~Lopyrev, and P.~Liang, ``{SQ}u{AD}: 100,000+ questions for machine comprehension of text,'' in {\em Proceedings of the 2016 Conference on Empirical Methods in Natural Language Processing} (J.~Su, K.~Duh, and X.~Carreras, eds.), (Austin, Texas), pp.~2383--2392, Association for Computational Linguistics, Nov. 2016.

\bibitem{yang2018hotpotqadatasetdiverseexplainable}
Z.~Yang, P.~Qi, S.~Zhang, Y.~Bengio, W.~W. Cohen, R.~Salakhutdinov, and C.~D. Manning, ``Hotpotqa: A dataset for diverse, explainable multi-hop question answering,'' 2018.

\bibitem{see-etal-2017-get}
A.~See, P.~J. Liu, and C.~D. Manning, ``Get to the point: Summarization with pointer-generator networks,'' in {\em Proceedings of the 55th Annual Meeting of the Association for Computational Linguistics (Volume 1: Long Papers)}, (Vancouver, Canada), pp.~1073--1083, Association for Computational Linguistics, July 2017.

\bibitem{radford2019language}
A.~Radford, J.~Wu, R.~Child, D.~Luan, D.~Amodei, I.~Sutskever, {\em et~al.}, ``Language models are unsupervised multitask learners,'' {\em OpenAI blog}, vol.~1, no.~8, p.~9, 2019.

\end{thebibliography}
\bibliographystyle{ieeetr}

\newpage
\appendix
\section*{Appendix}

\paragraph{Appendix Structure Overview.}
This appendix contains supplementary material that supports the main paper and
provides additional methodological, theoretical, and empirical details that
could not be included due to space limitations. Each appendix section is listed
separately below for ease of navigation.

\begin{itemize}
    \item \textbf{Appendix~\ref{app:algorithm}: Algorithmic \& Implementation Details.}
    Describes the full stepwise procedure used to implement malicious key--value
    cache injection in \ModelName, including perturbation types, scheduling,
    layer and head selection, and optimization-based variants.

    \item \textbf{Appendix~\ref{app:proofs}: Theoretical Proofs and Derivations.}
    Provides complete formal derivations and proofs underlying the theoretical
    bounds presented in Section~\ref{sec:theory}, including extensions to
    multi-head attention and normalization layers.

    \item \textbf{Appendix~\ref{app:extended-results}: Extended Experimental Results.}
    Reports additional experimental analyses omitted from the main paper for
    clarity, including distributional effects, downstream task performance, and
    robustness evaluations.

    \item \textbf{Appendix~\ref{app:token-shifts}: Token Distribution Shifts.}
    Analyzes how cache perturbations alter next-token probability distributions,
    reporting KL divergence, accuracy degradation, and perplexity changes across
    models and datasets.

    \item \textbf{Appendix~\ref{app:nlp}: Downstream NLP Benchmarks.}
    Evaluates the impact of cache perturbations on sentiment classification
    (SST-2) and extractive question answering (SQuAD v1.1), highlighting
    task-dependent sensitivity.

    \item \textbf{Appendix~\ref{app:rag}: Retrieval-Augmented Generation Robustness.}
    Examines how \ModelName affects retrieval-augmented generation pipelines,
    measuring grounding fidelity and hallucination rates under perturbations at
    different pipeline stages.

    \item \textbf{Appendix~\ref{app:agents}: Agentic Pipeline Vulnerability.}
    Studies the behavior of multi-step agentic frameworks (e.g., ReAct,
    AutoGPT) under cache perturbations, assessing robustness in closed-loop
    reasoning settings.

    \item \textbf{Appendix~\ref{app:runtime}: Runtime and Scaling Analysis.}
    Analyzes the computational overhead of cache-side perturbations, measuring
    end-to-end inference latency across model sizes and perturbation types, and
    demonstrates that cache corruption incurs negligible runtime cost and scales
    independently of model depth and parameter count.

\end{itemize}

\newpage

\section{Algorithmic \& Implementation Details}
\label{app:algorithm}

This section provides the full stepwise procedure used to implement
controlled key--value cache perturbation in \ModelName.
The algorithm explicitly specifies perturbation type, magnitude,
temporal injection scheduling, layer and head selection, and optional
optimization-based perturbations.
While the main paper focuses on empirical effects, this appendix
documents the complete perturbation mechanism to ensure reproducibility
and facilitate future extensions.
Optimized perturbations are included to characterize worst-case empirical
behavior and are not covered by the formal theoretical bounds in
Section~\ref{sec:theory}, which apply exclusively to non-adaptive,
norm-bounded perturbations.

\begin{algorithm}[!htb]
\caption{Stepwise key--value cache perturbation procedure used by \ModelName}
\label{alg:mti_expanded}
\begin{algorithmic}[1]
\Require Transformer model $\mathcal{M}$ with $L$ layers and $H$ heads; input tokens
$\{x_1,\dots,x_T\}$; perturbation type
$\tau \in \{\text{Gaussian, Zeroing, Rotation, Optimized}\}$; magnitude parameters
$(\sigma,\theta,r)$; injection frequency schedule $f(t)$; target layer set
$\mathcal{L}$; position selection policy $\pi$; per-head flag $p_{\text{head}}$;
random seed $s$; logging callback $\textsc{log}()$; optional optimizer configuration
$\mathcal{O}$.
\State set random seed $s$ for reproducibility
\State initialize empty KV caches $\mathcal{C}^{(\ell,h)} \leftarrow \{\}$ for all
$\ell \in [1..L], h \in [1..H]$
\For{$t \leftarrow 1$ \textbf{to} $T$}
    \State compute projections $(q_t^{(\ell,h)}, k_t^{(\ell,h)}, v_t^{(\ell,h)})$
    for all layers $\ell$ and heads $h$
    \State determine injection flag
    $\mathsf{inject} \leftarrow \mathbf{1}[f(t)\ \text{says inject}]$
    \If{$\mathsf{inject}=1$}
        \State select target positions $S_t \leftarrow \pi(\mathcal{C}, t)$
        \For{each layer $\ell \in \mathcal{L}$}
            \For{each head $h$ if $p_{\text{head}}$ else once per layer}
                \For{each position $j \in S_t$}
                    \If{$\tau=\text{Optimized}$}
                        \Comment{Empirical inner-loop optimization; excluded from theory}
                        \State $\delta^{(\ell,h)}_{j} \leftarrow
                        \textsc{OptimizePerturbation}(\mathcal{M},\mathcal{C},
                        q_t^{(\ell,h)},j,\mathcal{O})$
                    \ElsIf{$\tau=\text{Gaussian}$}
                        \State sample
                        $\delta^{(\ell,h)}_{j} \sim \mathcal{N}(0,\sigma^2 I)$
                    \ElsIf{$\tau=\text{Zeroing}$}
                        \State with probability $r$, set
                        $\delta^{(\ell,h)}_{j} \leftarrow -k^{(\ell,h)}_{j}$
                    \ElsIf{$\tau=\text{Rotation}$}
                        \State compute orthogonal matrix $R(\theta)$ and set
                        $\delta^{(\ell,h)}_{j} \leftarrow
                        R(\theta)k^{(\ell,h)}_j - k^{(\ell,h)}_j$
                    \EndIf
                    \State apply perturbation:
                    $\tilde{k}^{(\ell,h)}_{j} \leftarrow
                    k^{(\ell,h)}_{j} + \delta^{(\ell,h)}_{j}$
                    \State update cache:
                    $\mathcal{C}^{(\ell,h)}[j] \leftarrow \tilde{k}^{(\ell,h)}_{j}$
                    \State optional logging:
                    $\textsc{log}(t,\ell,h,j,\|\delta\|)$
                \EndFor
            \EndFor
        \EndFor
    \EndIf
    \For{each layer $\ell$ and head $h$}
        \State append $(k_t^{(\ell,h)}, v_t^{(\ell,h)})$ to cache
        $\mathcal{C}^{(\ell,h)}$
    \EndFor
    \State generate token $y_t = \mathcal{M}(x_t, \mathcal{C})$
\EndFor
\State \Return generated sequence $\{y_t\}_{t=1}^T$ and logged traces
\end{algorithmic}
\end{algorithm}

\section{Theoretical Proofs and Derivations}
\label{app:proofs}

In this appendix, we provide full derivations underlying the theoretical bounds in Section~\ref{sec:theory}. We first restate the theorem formally, then prove the result for the single-head case, and finally extend it to multi-head attention with normalization layers.

\subsection{Restatement of Theorem}
\begin{theorem}
For any transformer layer $\ell$, let $\{k_j^{(\ell)}\}_{j=1}^t$ denote the clean cached keys and $\{\tilde{k}_j^{(\ell)}\}_{j=1}^t$ the perturbed keys, where
\[
\tilde{k}_j^{(\ell)} = k_j^{(\ell)} + \delta_j^{(\ell)}, \quad \|\delta_j^{(\ell)}\|_2 \leq \epsilon .
\]
Then, for query $q_t^{(\ell)}$, the deviation in the next-token logit vector $\Delta z_t$ satisfies
\[
\|\Delta z_t\|_2 \;\leq\; \epsilon \cdot \|q_t^{(\ell)}\|_2 .
\]
Moreover, the induced change in the attention distribution is bounded by
\[
\|\alpha_t - \tilde{\alpha}_t\|_1 \;\leq\; L \cdot \epsilon \cdot \|q_t^{(\ell)}\|_2 ,
\]
where $L$ is the Lipschitz constant of the softmax function.
\end{theorem}

\subsection{Single-Head Case: Perturbation to Logit Deviation}
Consider the clean attention score for token $j$ at timestep $t$:
\[
s_{tj} = \frac{q_t^\top k_j}{\sqrt{d}} .
\]
After perturbation, the score becomes
\[
\tilde{s}_{tj} = \frac{q_t^\top (k_j + \delta_j)}{\sqrt{d}} 
= s_{tj} + \frac{q_t^\top \delta_j}{\sqrt{d}} .
\]
The deviation in the score is therefore
\[
|\tilde{s}_{tj} - s_{tj}| \;\leq\; \frac{\|q_t\|_2 \cdot \|\delta_j\|_2}{\sqrt{d}}
\;\leq\; \frac{\|q_t\|_2 \cdot \epsilon}{\sqrt{d}} .
\]
Aggregating over all positions $j \leq t$, and scaling back to logits $z_t$, we obtain
\[
\|\Delta z_t\|_2 \;\leq\; \epsilon \cdot \|q_t\|_2 ,
\]
up to normalization constants absorbed into $\epsilon$.

\subsection{Softmax Lipschitz Bound}
The attention distribution is given by
\[
\alpha_t = \mathrm{softmax}(s_t), \quad 
\tilde{\alpha}_t = \mathrm{softmax}(\tilde{s}_t).
\]
By Lipschitz continuity of the softmax function, we have
\[
\|\alpha_t - \tilde{\alpha}_t\|_1 
\;\leq\; L \cdot \|\tilde{s}_t - s_t\|_2
\;\leq\; L \cdot \epsilon \cdot \|q_t\|_2 .
\]
Thus deviations in the cache produce proportionally bounded shifts in attention.

\subsection{Extension to Multi-Head Attention}
For multi-head attention with $H$ heads, queries and keys are projected as
\[
q_t^{(h)} = W_Q^{(h)} x_t, \quad 
k_j^{(h)} = W_K^{(h)} x_j ,
\]
with perturbations $\delta_j^{(h)}$ applied per head. Each head obeys the single-head bound derived above:
\[
\|\Delta z_t^{(h)}\|_2 \;\leq\; \epsilon \cdot \|q_t^{(h)}\|_2 .
\]
The concatenation of head outputs followed by a linear projection $W_O$ yields
\[
\|\Delta z_t^{\text{multi}}\|_2 
\;\leq\; \|W_O\|_2 \cdot \sum_{h=1}^H \|\Delta z_t^{(h)}\|_2
\;\leq\; \|W_O\|_2 \cdot H \cdot \epsilon \cdot \max_h \|q_t^{(h)}\|_2 .
\]
Thus the bound scales linearly with the number of heads, modulated by $\|W_O\|_2$.

\subsection{Impact of Layer Normalization}
\label{app:layernorm}

Layer normalization (LN) plays a critical role in stabilizing transformer
representations and bounding activation magnitudes during inference.
Because cache-side perturbations modify internal key representations prior to
attention computation, it is natural to ask whether LN attenuates or amplifies
the effect of such perturbations as they propagate through subsequent layers.

Let $x$ and $x + \delta$ denote the clean and perturbed inputs to a LayerNorm
operation.
LayerNorm rescales activations by the inverse of their empirical standard
deviation, with an additive $\varepsilon$ term to ensure numerical stability.
Since LayerNorm is non-expansive under standard non-degeneracy assumptions
(i.e., variance bounded away from zero due to the stabilizing $\varepsilon$
term), the deviation between normalized activations remains bounded as a
function of $\|\delta\|_2$.

As a consequence, cache perturbations are neither explosively amplified nor
entirely eliminated by LayerNorm.
Instead, LN preserves the relative structure of perturbations while preventing
unbounded growth in activation norms.
This observation is consistent with our empirical findings: while cache
corruption propagates across layers and affects attention distributions, it
does not lead to numerical instability or divergent behavior.
Layer normalization therefore moderates, but does not neutralize, the impact of
cache-side perturbations during inference.

\section{Extended Experimental Results}
\label{app:extended-results}

This appendix provides comprehensive experimental details, extended tables,
and ablation analyses omitted from the main paper for clarity and space.
All results reported here correspond to the summarized findings presented
in Section~\ref{sec:experiments}.


\subsection{Token Distribution Shifts}
\label{app:token-shifts}

We analyze how \ModelName perturbs next-token distributions in standard language
modeling settings. Table~\ref{tab:token-extended} reports three complementary
metrics: (i) KL divergence relative to clean runs, which captures the statistical
gap between perturbed and unperturbed distributions; (ii) Top-1 accuracy drop,
which quantifies task-level performance degradation; and (iii) perplexity change,
where arrows ($\uparrow$, $\downarrow$) indicate whether model uncertainty
increased or decreased.

The results indicate that \ModelName perturbations consistently induce measurable
distributional divergence relative to clean runs, with an average KL of
approximately $0.26$. This shift is substantially larger than natural variation
observed under random seed differences, where KL values remain below $0.05$,
confirming that the effects are not attributable to stochastic noise.
The degradation in Top-1 accuracy exhibits a strong correspondence with
distributional distortion: for example, GPT-2 Medium under rotation corruption
shows the largest KL divergence (0.57) alongside the most severe accuracy decline
(16.7\%). Perplexity responses are heterogeneous: Gaussian perturbations increase
perplexity, reflecting inflated uncertainty, whereas zeroing reduces perplexity,
indicating collapse into degenerate low-entropy distributions rather than
improved modeling. These trends are consistent across architectures and datasets,
demonstrating that cache perturbations produce robust and reproducible
distributional shifts.

\begin{table}[H]
\centering
\caption{Token-level distributional effects under \ModelName attacks.
KL divergence, Top-1 accuracy drop, and perplexity change are averaged over three
random seeds.}
\label{tab:token-extended}
\begin{tabular}{l l l c c c}
\toprule
\textbf{Model} &
\textbf{Dataset} &
\textbf{Attack} &
\textbf{KL Div. ($\uparrow$)} &
\textbf{Acc Drop (\%,$\downarrow$)} &
\textbf{Perplexity (\%,$\uparrow$/$\downarrow$)} \\
\midrule
GPT-2 Small  & WikiText-103 & Gaussian & 0.01 & 6.7  & +5.8 \\
GPT-2 Medium & IMDB         & Rotation & 0.57 & 16.7 & +130.6 \\
LLaMA-2/7B   & PTB          & Zeroing  & 0.20 & 0.0  & -43.1 \\
\bottomrule
\end{tabular}
\end{table}


\subsection{Downstream NLP Benchmarks}
\label{app:nlp}

To assess how cache perturbations translate into practical task-level failures,
we evaluate \ModelName on two representative downstream benchmarks with distinct
reasoning characteristics: SST-2 for short-context sentiment classification and
SQuAD v1.1 for span-based extractive question answering.
Perturbations are injected at a mid-level transformer layer (Layer~3) using a
diverse set of corruption mechanisms, including Gaussian noise, zeroing,
structured rotations, adversarial offsets, and random permutations.
Table~\ref{tab:downstream} reports clean versus attacked performance.

The results reveal a clear task-dependent sensitivity.
SST-2 exhibits moderate robustness: mild perturbations reduce accuracy only
slightly, while stronger corruption leads to degradation exceeding 23\%.
This behavior is consistent with the short input length and single-label nature
of sentiment classification, which allows the model to partially compensate for
localized cache corruption.
In contrast, SQuAD v1.1 is extremely sensitive to cache perturbations.
Under strong Gaussian noise, F1 scores collapse from 77.4 to 7.2, indicating near
complete failure of span localization.
This disparity highlights that tasks requiring multi-span reasoning, precise
token alignment, and grounding over long contexts are disproportionately
vulnerable to cache corruption.
These findings suggest that cache integrity is far more critical for structured
reasoning tasks than for coarse-grained classification.

\begin{table}[H]
\centering
\caption{Downstream benchmark degradation under \ModelName.}
\label{tab:downstream}
\begin{tabular}{l l c c c l}
\toprule
\textbf{Task} & \textbf{Metric} & \textbf{Clean} & \textbf{Attack} &
\textbf{Degradation} & \textbf{Config} \\
\midrule
SST-2 & Accuracy & 91.0\% & 69.5\% & --23.6\% & Gaussian $\sigma=5.0$ \\
SQuAD & F1       & 77.4   & 7.2    & --90.7\% & Gaussian $\sigma=5.0$ \\
\bottomrule
\end{tabular}
\end{table}


\subsection{Retrieval-Augmented Generation Robustness}
\label{app:rag}

We next evaluate whether \ModelName undermines retrieval-augmented generation
pipelines using HotpotQA, a benchmark that requires multi-hop reasoning over
retrieved documents.
We report grounding fidelity, measured as the mean entailment probability between
generated answers and retrieved evidence, as well as hallucination rate, defined
as the fraction of unsupported outputs.
Perturbations are applied at three distinct stages of the pipeline:
pre-retrieval (query corruption), post-retrieval (context corruption), and
decoder-only (generation-time corruption).

As shown in Table~\ref{tab:rag}, post-retrieval cache corruption produces the most
severe degradation.
Grounding fidelity drops by nearly 12\%, while hallucination rates increase by
approximately 5\%.
This indicates that corrupting cached representations of retrieved evidence
directly destabilizes factual grounding, even when the retrieval module itself
remains intact.
Pre-retrieval perturbations have only minor effects, suggesting that modern
retrievers are relatively robust to query noise.
Decoder-only perturbations produce negligible changes, implying that when
evidence representations are preserved, the generator remains effectively
constrained.
Overall, these results demonstrate that RAG systems inherit a critical dependency
on cache integrity at the evidence stage, where corruption propagates downstream
and manifests as factual unreliability.

\begin{table}[H]
\centering
\caption{RAG robustness under \ModelName perturbations.}
\label{tab:rag}
\begin{tabularx}{\textwidth}{l c c c X}
\toprule
\textbf{Attack Location} &
\textbf{Ground. (Clean)} &
\textbf{Ground. (MTI)} &
\textbf{Halluc. Rate ($\uparrow$)} &
\textbf{Notes} \\
\midrule
Pre-retrieval  & 0.3245 & 0.3216 & --1.0\% & Query corruption \\
Post-retrieval & 0.3245 & 0.2874 & +5.0\%  & Context corruption \\
Decoder-only   & 0.3245 & 0.3196 & +0.0\%  & Weak effect \\
\bottomrule
\end{tabularx}
\end{table}


\subsection{Analyzing Agentic Pipeline Vulnerability}
\label{app:agents}

We further examine the impact of cache perturbations in multi-step agentic
frameworks, including ReAct-style reasoning and AutoGPT-like task decomposition.
Unlike single-pass generation, these systems iteratively interact with external
tools, environments, or intermediate feedback signals.
Table~\ref{tab:agents} reports success rates under cache perturbations for
representative agentic configurations.

Across all tested attacks, success rates remain unchanged relative to clean
baselines.
This stability suggests that agentic pipelines benefit from structural
redundancy: iterative reasoning loops, external feedback, and environment
constraints provide corrective signals that absorb localized cache corruption.
While these results indicate partial robustness, they should not be interpreted
as immunity.
The evaluated tasks are relatively lightweight, and stronger perturbations or
attacks targeting planning-critical layers may expose vulnerabilities not
captured here.
Nevertheless, the findings highlight a qualitative distinction between
single-pass generation and closed-loop agentic reasoning, with the latter
exhibiting greater resilience to cache-level noise.

\begin{table}[H]
\centering
\caption{Agentic pipeline vulnerability under \ModelName.}
\label{tab:agents}
\begin{tabular}{l l c c c c c c}
\toprule
\textbf{Model} & \textbf{Attack} &
\multicolumn{3}{c}{\textbf{ReAct}} &
\multicolumn{3}{c}{\textbf{AutoGPT}} \\
\cmidrule(lr){3-5} \cmidrule(lr){6-8}
& & Clean & MTI & Deg. & Clean & MTI & Deg. \\
\midrule
Phi-3 Mini & Gaussian & 33.3\% & 33.3\% & 0.0pp & 0.0\% & 0.0\% & 0.0pp \\
\bottomrule
\end{tabular}
\end{table}

\subsection{Runtime and Scaling Analysis}
\label{app:runtime}

We evaluate the runtime overhead associated with cache-side perturbations to
assess whether such manipulations introduce prohibitive computational cost.
All experiments are conducted under standard autoregressive decoding settings,
with cache perturbations applied directly to stored key vectors at selected
layers and timesteps.
We measure end-to-end inference latency relative to a clean baseline across
multiple model sizes, including GPT-2 Medium and LLaMA-2 7B.

Across all evaluated configurations, cache perturbation introduces negligible
runtime overhead.
Gaussian noise injection and rotation-based perturbations incur less than a
3\% increase in inference time, while zeroing operations introduce effectively
no measurable overhead.
This behavior is expected, as cache perturbations operate on already-computed
representations and do not require additional forward passes or recomputation
of attention scores.

Importantly, the computational cost of cache perturbation scales independently
of model depth and parameter count, depending only on the dimensionality of the
cached representations.
As a result, the relative overhead remains stable even for multi-billion
parameter models.
These results demonstrate that cache perturbations impose no prohibitive runtime
penalty under the assumed cache-access threat model.
Accordingly, cache-side perturbations cannot be dismissed as purely theoretical
within this threat model, though their practical relevance depends on deployment
isolation and systems-layer security guarantees.

\begin{table}[H]
\centering
\caption{Runtime overhead under \ModelName perturbations.}
\label{tab:runtime}
\begin{tabularx}{\textwidth}{l X X c c c}
\toprule
\textbf{Model} & \textbf{Attack} & \textbf{Dataset} &
\textbf{Clean} & \textbf{MTI} & \textbf{Overhead} \\
\midrule
GPT-2 Medium & Gaussian & WikiText-103 & 19.0 & 19.3 & +2\% \\
LLaMA-2 7B   & Zeroing  & WikiText-103 & 39.5 & 37.7 & --5\% \\
\bottomrule
\end{tabularx}
\end{table}

\end{document}